\documentclass[trackchanges]{aastex701}

\usepackage{amsmath}
\begin{document}

\title{On the dual nature of atmospheric escape}

\author[orcid=0000-0001-6425-9415]{Darius Modirrousta-Galian}
\affiliation{Tsung‐Dao Lee Institute, Shanghai Jiao Tong University, Shanghai, China}
\affiliation{School of Physics and Astronomy, Shanghai Jiao Tong University, Shanghai, China}
\email[show]{modirrousta-galian@sjtu.edu.cn}  

\author[orcid=0000-0002-4785-2273]{Jun Korenaga} 
\affiliation{Department of Earth and Planetary Sciences, Yale University, New Haven, CT, USA}
\email{jun.korenaga@yale.edu}

\begin{abstract}
    Planetary atmospheres cannot remain hydrostatic at all altitudes because they approach finite density at infinite radius, implying infinite mass. Classical treatments address this in two directions: either retain a hydrostatic structure while allowing particles in the high-velocity tail to decouple and escape in a Jeans-type manner, or promote the gas to a continuum outflow to obtain a transonic Parker-type solution. The usual criterion compares the local mean free path to the sonic point radius. If the mean free path is shorter, the atmosphere is hydrostatic with an imposed Jeans escape flux; if it is longer, the gas is hydrodynamic with Jeans escape neglected. Here, we show that hydrogen-rich atmospheres do not separate cleanly into hydrodynamic and Jeans-escape regimes. At any radius, some particles still collide and behave as a fluid, while others have already experienced their last collision and move collisionlessly on ballistic trajectories. The relative importance of these two behaviors changes smoothly with radius rather than switching at a single boundary. The hydrodynamic channel accelerates and passes through a sonic point, whereas the collisionless channel decelerates under gravity and grows with altitude, removing mass and momentum from the collisional flow. As the collisionless component grows, the bulk flow speed reaches a maximum and then decelerates thereafter, producing profiles similar to Parker breeze solutions even though escape is carried by the collisionless channel. This two-channel framework provides a first step toward a self-consistent treatment that unifies hydrodynamics and kinetics in atmospheric loss models.
\end{abstract}


\section{Introduction} 

In the 1950s, David Bates discovered that planetary thermospheres have a temperature inversion, which is well approximated as a conductive temperature profile driven by X-ray and ultraviolet absorption \citep{Bates1951,Bates1959}. However, applying his static conduction model to hydrogen-rich atmospheres yielded implausibly high temperatures of the order $10^{4}{-}10^{5}~{\rm K}$ \citep{Gross1972,Horedt1982}. Several years prior, Eugene Parker suggested that the solar atmosphere had to be hydrodynamic \citep{Parker1958,Parker1964,Parker1964b} to prevent its density from following the barometric formula, which erroneously predicts non-zero density at infinite radii \citep[][pages~147$-$151]{walker1977}. Indeed, the Lunik~III and Venus~I missions confirmed the presence of solar winds by detecting the flux of charged particles emanating from the Sun, validating Parker's theoretical insights. Expanding upon Parker's theory, \citet{Sekiya1980,Sekiya1981} suggested that irradiated hydrogen-rich planetary atmospheres cannot sustain hydrostatic equilibrium and they proposed a hydrodynamic outflow model, which provided a more realistic temperature profile. 

There has since been a long-standing effort to elaborate the model and incorporate additional effects. For example, \citet{Watson1981} considered conduction and advection concurrently, \citet{Zahnle1986} evaluated the effects of mass fractionation, and \citet{Chassefiere1996} accounted for solar wind ram pressure. The discovery of exoplanets in the 1990s \citep{Wolszczan1992,Mayor1995} and early 2000s \citep{Udalski2002,Bond2004}, characterized by their elevated temperatures and highly-irradiated environments, provided a platform for refining photoevaporation models. \citet{Garcia2007} and \citet{Murray2009} incorporated a chemical, ionization, and recombination model, \citet{Kubyshkina2018b} investigated the effects of high equilibrium temperatures, \citet{Owen2019b} assessed the importance of nonzero planetary magnetic fields on mass outflow, and \citet{Modirrousta2024} showed that momentum diffusion between hydrogen and heavier species can significantly reduce mass loss rates by several orders of magnitude.

Whereas these developments have advanced our understanding of atmospheric escape, a key aspect remains poorly understood: how a collisional, continuum outflow connects to a collisionless kinetic escape at large radii. Hydrodynamic analyses treat the atmosphere as a continuous, collisional fluid and accurately describe the deep regions where particle collisions are frequent \citep{Parker1964,Parker1964b,Hunten1982}. However, this framework is inadequate at higher altitudes, where the gas becomes rarefied, collisions are infrequent, and the flow gradually transitions to a collisionless, ballistic regime. Atomistic methods such as Direct Simulation Monte Carlo (DSMC) are well suited for modeling this upper, collisionless domain \citep{Volkov2011a,Volkov2013a}. Whereas the Boltzmann equation can in principle capture both collisional and collisionless behavior, DSMC simulations rely on the assumption of molecular chaos, which neglects correlations between particle velocities \citep{Bird1998,Oran1998}. Such correlations become important for dense-gas effects and collective dynamics beyond the Boltzmann description. The alternative is to account for the degree of correlation between colliding particles by tracking their collisional history, but this lies outside current computational capabilities. Even with this simplification, DSMC calculations are computationally demanding, and solving the Boltzmann equation across an atmosphere that spans many orders of magnitude in density is not currently feasible. In practice, most studies adopt one approach or the other. In cases where a hybrid strategy is attempted, the coupling is usually implemented by matching the hydrodynamic solution to a DSMC calculation at a prescribed transition radius rather than by solving the coupled problem self-consistently: a hydrodynamic model is integrated outward from the deep atmosphere up to that radius, generally the exobase, i.e., where the mean free path equals the pressure scale height or the sonic point radius, and a DSMC calculation is then initialized at that boundary to represent the collisionless exosphere \citep{Kislyakova2019,Kislyakova2020}. This procedure has two important limitations. First, the hydrodynamic solution already fixes the density, velocity, and temperature, so the DSMC calculation does not determine the escape rate independently. Second, the predicted structure of the flow depends on the arbitrarily chosen location of the transition boundary, introducing a bias that is difficult to quantify. Thus, DSMC simulations cannot model collisional flow at low radii, hydrodynamic models cannot capture collisionless behavior at large radii, and the transition region, where the gas is neither fully collisional nor fully ballistic, remains poorly represented by existing approaches.

This study develops a hybrid hydrodynamic framework that keeps a collisional outflow while explicitly allowing a fraction of particles to decouple and escape ballistically. The key change is that mass and momentum are not conserved within a single fluid channel: they are transferred continuously from a collisional channel to a collisionless channel as the gas rarefies. In this picture, the collisional component still admits a transonic Parker-type solution, while the flux-weighted bulk (i.e., the combination of the two channels) velocity can develop a maximum and then decline at larger radii as the collisionless component grows. This provides a route to resolve the long-standing ambiguity in how Parker-type transonic winds connect to Jeans-type escape, without imposing an arbitrary transition radius. We quantify this behavior by introducing a decoupling fraction that controls the local transfer between channels and by defining a quasi-sonic point as the radius where the bulk speed reaches its maximum. The structure of this paper is as follows. In Section~\ref{sec:methods}, we define the decoupling fraction and derive modified continuity and momentum equations with explicit sink terms for the collisional channel. In Section~\ref{sec:results}, we present representative velocity and density profiles and use them to identify the quasi-sonic point in the bulk flow. In Section~\ref{sec:discussion}, we address the observability of atmospheric escape under a breeze-like bulk profile and clarify the limitations and intended scope of the model. In Section~\ref{sec:conclusion}, we summarize the main conclusions.

\section{Methods}
\label{sec:methods}
\subsection{Modified conservation equations and the sonic point} 

To incorporate kinetic effects into a hydrodynamic model, we first define a metric that measures the fraction of particles that decouple from the hydrodynamic flow and escape ballistically without further collisions. We refer to $\phi(r)$ as the decoupling fraction. It is defined as the local fraction of particles at radius $r$ that satisfy three requirements simultaneously: they move radially outward, they undergo no further collisions after leaving $r$, and they are unbound in the planet frame. The first requirement keeps only particles whose velocities point outward, that is, those in the outward half of the local velocity distribution. The second is the fraction of particles that experience no further collisions along their outward trajectory. The third selects the portion of the comoving velocity distribution whose excess speed above the bulk flow is sufficient for escape. The decoupling fraction $\phi(r)$ can therefore be written as the product of an outward-direction fraction, a survival probability $P_{\rm sur}$ (i.e., no further collisions), and an unbound-tail fraction $P_{\rm esc}$,
\begin{equation}
    \phi(r)=\frac{1}{2} P_{\rm sur}(r) P_{\rm esc}(r),
\label{eq:phi_1}
\end{equation}
where,
\begin{equation}
    P_{\rm sur}(r) = \exp\left[-\int_{r}^{\infty} \sqrt{2}n(r)\sigma{\rm d}r\right],
\label{eq:P_sur}
\end{equation}
and
\begin{equation}
    P_{\rm esc}(r) = \int_{u_0(r)}^{\infty}\left[\frac{\bar\mu}{2\pi k_{\rm B}T(r)}\right]^{3/2} 4\pi u^{2} \exp\left[-\frac{\bar\mu u^{2}}{2k_{\rm B}T(r)}\right]{\rm d}u,
\label{eq:P_esc}
\end{equation}
with $u_0(r){=}\max\left[0,v_{\rm esc}(r){-}v(r)\right]$. Probability $P_{\rm sur}(r){\to}1$ when the density $n$ above $r$ becomes negligible and there are effectively no particles left to collide with along the outward path. By construction, $P_{\rm esc}(r){\to}1$ when $u_0(r){\to}0$ because the Maxwellian is normalized over $u{\in}[0,\infty)$. Here, $v(r)$ is the bulk radial velocity of the collisional flow and $v_{\rm esc}(r){=}\sqrt{2GM/r}$ is the minimum escape speed. The integration variable $u$ denotes the particle’s random thermal speed in the local fluid rest frame. The bulk wind provides an additional outward drift in the planet frame. Thus, $u_{0}$ is the minimum additional thermal speed, in addition to the bulk drift, required for a particle to be unbound. Because thermal motions are isotropic, we include only the outward moving half of the thermal velocity distribution when estimating the escaping population, i.e., the prefactor of one half. Equation~\ref{eq:P_sur} represents the survival probability that a particle initially at radius $r$ reaches infinity without undergoing another collision \citep[][page\,153]{walker1977}. This is obtained by integrating the collisional attenuation $\sqrt{2}n\sigma$ along its outward trajectory, where $n(r)$ is the particle number density, $\sigma$ is the collisional cross section, and the factor of $\sqrt{2}$ accounts for the relative speed of colliding particles in a Maxwellian gas (see Appendix~\ref{app:mfp}). Equation~\ref{eq:P_esc} accounts for the fraction of the local Maxwellian distribution that exceeds the threshold excess speed $u_0$. 

The local rate at which particles decouple from the collisional channel and enter the collisionless channel can then be expressed as the product of three factors: the number density of escaping particles at radius $r$, $n\phi$; the inverse mean free path, $l_{\rm mfp}^{-1}(r){=}\sqrt{2}n(r)\sigma$, where the probability that a particle collides while traversing a distance ${\rm d}r$ is $1{-}\exp[-{\rm d}r/l_{\rm mfp}(r)]{\approx}{\rm d}r/l_{\rm mfp}(r)$ for ${\rm d}r{\ll} l_{\rm mfp}(r)$, so multiplying by $l_{\rm mfp}^{-1}$ converts a path length into a collision probability; and the outward radial speed with which such particles leave the collisional fluid, $v{+}\bar{v}_{\rm esc}$, where $\bar{v}_{\rm esc}{=}F_1(v)v_{\rm esc}$ and $F_1(v)$ is the dimensionless mean relative extra speed in the comoving frame, normalized by $v_{\rm esc}$,
\begin{equation}
    F_{1}(v)=\frac{1}{v_{\rm esc}(r)}\frac{\int_{u_0(r)}^{\infty} f(u) u~{\rm d}u }{\int_{u_0(r)}^{\infty} f(u) ~{\rm d}u },
\label{eq:F1_definition}
\end{equation}
with
\begin{equation}
    f(u) =\left[\frac{\bar{\mu}}{2\pi k_{\rm B}T(r)}\right]^{\frac{3}{2}}4 \pi u^2 \exp{\left[-\frac{\bar{\mu}u^{2}}{2k_{\rm B}T(r)}\right]}.
\end{equation}
Here, $\bar{v}_{\rm esc}(r)$ denotes the mean velocity of particles that instantaneously decouple from the flow and escape at radius $r$. Because escape occurs with a distribution of velocities above the minimum escape velocity, $\bar{v}_{\rm esc}$ is defined as the Maxwellian-weighted mean over the unbound tail, i.e., over comoving thermal velocities satisfying $u{>}u_0(r)$.

The product of the three factors, $\sqrt{2}n^{2}\sigma \phi (v{+}\bar v_{\rm esc})$, gives the volumetric removal rate of particles from the collisional flow at radius $r$ (See Appendix~\ref{app:mass_sink} for further discussion). The mass continuity equation therefore reads,
\begin{equation}
    \frac{1}{r^2}\frac{{\rm d}}{{\rm d}r}\left(n v r^{2}\right) = - \sqrt{2}n^{2}\sigma \phi (v+\bar v_{\rm esc}),
\label{eq:to_integrate}
\end{equation}
which can be expressed as,
\begin{equation}
    \frac{1}{n}\frac{{\rm d}n}{{\rm d}r} + \frac{1}{v}\frac{{\rm d}v}{{\rm d}r} + \frac{2}{r} = - \sqrt{2}n \sigma \phi\left(1+\frac{\bar{v}_{\rm esc}}{v}\right).
\label{eq:mass_conservation}
\end{equation}

The mass removed per unit volume per unit time is $\sqrt{2}n^{2} \bar{\mu} \sigma \phi (v{+}\bar{v}_{\rm esc})$, and those particles leave the fluid with relative radial speed $\bar{v}_{\rm esc}{=}F_1(v)v_{\rm esc}$. Hence the momentum sink per unit volume per unit time acting on the collisional fluid is $\sqrt{2}n^{2} \bar{\mu} \sigma \phi (v{+}\bar{v}_{\rm esc})\bar{v}_{\rm esc}$. Here $\bar{\mu}$ is the particle mass, taken as constant, introduced only to convert the number removal rate into mass and momentum sink terms. The momentum conservation equation then reads, 
\begin{equation}
    \frac{1}{n \bar{\mu}}\frac{{\rm d}P}{{\rm d}r} = -v\frac{{\rm d}v}{{\rm d}r} - \frac{GM}{r^{2}} - \sqrt{2}n\sigma \phi \bar{v}_{\rm esc}^{2}\left(1+\frac{v}{\bar{v}_{\rm esc}}\right),
\label{eq:mom_conservation}
\end{equation}
which reduces to the standard Parker result in the limit $\phi{\to}0$.

The last equation we introduce is the ideal gas law, which serves as our equation of state. Differentiating it gives,
\begin{equation}
    \frac{1}{n \bar{\mu}}\frac{{\rm d}P}{{\rm d}r} = \frac{k_{\rm B}T}{\bar{\mu}} \left(\frac{1}{n}\frac{{\rm d}n}{{\rm d}r} + \frac{1}{T}\frac{{\rm d}T}{{\rm d}r}\right).
\label{eq:EOS1}
\end{equation}
If the gas follows a polytropic temperature profile, Equation~\ref{eq:EOS1} simplifies to,
\begin{equation}
\frac{1}{n \bar{\mu}} \frac{{\rm d}P}{{\rm d}r} = \frac{\gamma k_{\rm B} T}{\bar{\mu}} \frac{1}{n} \frac{{\rm d}n}{{\rm d}r} \equiv c^{2}_{\rm s} \frac{1}{n} \frac{{\rm d}n}{{\rm d}r},
\label{eq:EOS2}
\end{equation}
with $c_{\rm s}$ and $\gamma$ denoting the sound speed and polytropic index. When $\gamma$ is identified with the heat capacity ratio, $\gamma {=} 1$ corresponds to an isothermal gas and $\gamma {\simeq} 1.1{-}1.67$ to adiabatic profiles. In the astrophysical literature, the polytropic index is defined by $P{\propto}n^{1+1/m}$, with $m$ as the index \citep[e.g.,][Chapter 4]{Chandrasekhar1939}. Here, we adopt $P{\propto}n^\gamma$, using $\gamma$ as the exponent for notational consistency with the heat capacity ratio of an ideal gas. We emphasize that the polytropic relation in Equation~\ref{eq:EOS2} is not intended as a complete description of thermospheric energy balance. Photoionization heating, line cooling, chemistry, and temperature inversions can all produce non-polytropic temperature profiles in real atmospheres. Our aim here is to keep the thermodynamic framework straightforward and tunable so that the impact of collisionality, through $\phi(r)$, can be identified clearly. To this end, $\gamma$ serves as a control parameter for the background temperature structure, and the key conclusions concern how the collisional-to-collisionless transfer alters the bulk flow for a given thermal profile.

Combining Equation~\ref{eq:mom_conservation} with \ref{eq:EOS2}, and solving for the density gradient, yields,
\begin{equation}
\frac{1}{n} \frac{{\rm d}n}{{\rm d}r} = -\frac{v}{c^{2}_{\rm s}}\frac{{\rm d}v}{{\rm d}r} - \frac{GM}{c^{2}_{\rm s}r^{2}} - \sqrt{2}n\sigma \phi \frac{\bar{v}_{\rm esc}^{2}}{c^{2}_{\rm s}}\left(1+\frac{v}{\bar{v}_{\rm esc}}\right),
\label{eq:hybrid_equation}
\end{equation}
which can be substituted into Equation~\ref{eq:mass_conservation}, and solved for the velocity gradient,
\begin{equation}
    \frac{{\rm d}v}{{\rm d}r} = \frac{2c_{\rm s}^2v}{r^{2}}\frac{\left[\sqrt{2} n \sigma \phi\frac{\left(v+\bar{v}_{\rm esc}\right)\left(v\bar{v}_{\rm esc}-c_{\rm s}^2\right)}{2vc_{\rm s}^2}\right]r^{2}-r+\frac{GM}{2 c_{\rm s}^2}}{c_{\rm s}^2-v^{2}}.
\label{eq:dv_dr}
\end{equation}
For a sonic point to exist, the flow must pass smoothly through $v{=}c_{\rm s,s}$ (i.e., the speed of sound at the sonic point). At this radius, $r{=}R_{\rm s}$, the denominator of the velocity gradient vanishes, so a finite solution is possible only if the numerator also vanishes. This regularity condition defines the sonic point. Setting the numerator to zero and recognizing that at the sonic point, $\bar{v}_{\rm esc}{=}F_1(c_{\rm s,s})v_{\rm esc}{=}F_1(c_{\rm s,s})\sqrt{2GM/R_{\rm s}}$, we obtain,
\begin{equation}
    \frac{\phi_{\rm s}\left(\frac{l_{\rm mfp,s}}{R_{\rm s,0}}\right)^{-1}}{2\left[2F_1(c_{\rm s,s})^{2}\phi_{\rm s}\left(\frac{l_{\rm mfp,s}}{R_{\rm s,0}}\right)^{-1}-1\right]}\left(\frac{R_{\rm s}}{R_{\rm s,0}}\right)^{2} - \left(\frac{R_{\rm s}}{R_{\rm s,0}}\right) - \frac{1}{2F_1(c_{\rm s,s})^{2}\phi_{\rm s}\left(\frac{l_{\rm mfp,s}}{R_{\rm s,0}}\right)^{-1}-1} = 0,
\label{eq:Rs_R0}
\end{equation}
where $R_{\rm s,0}{=}GM/(2c_{\rm s,s}^{2})$ is the classical Parker sonic point radius and $l_{\rm mfp,s}{=}\left(\sqrt{2}n_{\rm s}\sigma\right)^{-1}$ is the mean free path at the sonic point. Evaluating Equation~\ref{eq:F1_definition} for $F_{1}(c_{\rm s,s})$, we obtain,
\begin{equation}
\begin{split}
    F_{1}(c_{\rm s,s}) {=}  
\begin{cases}
    \frac{\frac{\gamma}{2}\zeta^{2}+1}{\gamma\zeta+\sqrt{\frac{\pi\gamma}{2}}\exp{\left(\frac{\gamma}{2}\zeta^{2}\right)}\operatorname{erfc}\left(\sqrt{\frac{\gamma}{2}}\zeta\right)} \sqrt{\frac{R_{\rm s}}{R_{\rm s,0}}},& \text{if } R_{\rm s}/R_{\rm s,0}{\leq}4\\ \\
    \sqrt{\frac{2}{\pi \gamma}}\sqrt{\frac{R_{\rm s}}{R_{\rm s,0}}}.              & \text{otherwise}
\end{cases},
\end{split}
\label{eq:F1_final}
\end{equation}
where $\zeta {=} 2\left(R_{\rm s}/R_{\rm s,0}\right)^{-\frac{1}{2}}{-}1$. As we show later in the manuscript, the ratio $R_{\rm s}/R_{\rm s,0}$ is bounded from above by ${\sim}1.3$. Consequently, only the first branch of Equation~\ref{eq:F1_final} is physical.

The last unknown is $\phi_{\rm s}$, which is found by evaluating Equation~\ref{eq:phi_1} at the sonic point, yielding,
\begin{equation}
    \phi_{\rm s}=\frac{1}{2} \exp\left[-\int_{R_{\rm s}}^{\infty} \sqrt{2} n\sigma{\rm d}r\right] \left[\sqrt{\frac{2\gamma}{\pi}}\zeta\exp{\left(-\frac{\gamma}{2}\zeta^{2}\right)} + \operatorname{erfc}{\left(\sqrt{\frac{\gamma}{2}}\zeta \right)} \right].
\label{eq:phi_2}
\end{equation}
We now introduce the local density scale height at the sonic point,
\begin{equation}
    H_{\rm s} \equiv -\left(\left.\frac{{\rm d}\ln n}{{\rm d}r}\right|_{R_{\rm s}}\right)^{-1}.
\end{equation}
A first-order Taylor expansion about $R_{\rm s}$ yields,
\begin{equation}
    \ln n \approx \ln n_{\rm s} + (r-R_{\rm s})\left.\frac{{\rm d}\ln n}{{\rm d}r}\right|_{R_{\rm s}} + \mathcal{O}\left[(r-R_{\rm s})^2\right].
\end{equation}
Keeping only the first order terms, using the definition of $H_{\rm s}$, and exponentiating, yields,
\begin{equation}
    n =  n_{\rm s} \exp{\left(-\frac{r-R_{\rm s}}{H_{\rm s}}\right)}.
\label{eq:n_r}
\end{equation}
Inserting Equation~\ref{eq:n_r} into \ref{eq:phi_2}, and recognizing $l_{\rm mfp,s}{=}\left(\sqrt{2}\sigma n_{\rm s}\right)^{-1}$ and $H_{\rm s}{=}1/\left(2\gamma\right)\left(R_{\rm s}/R_{\rm s,0}\right)^{2}R_{\rm s,0}$, we obtain,
\begin{equation}
    \phi_{\rm s} {=} \frac{1}{2} \exp\left[{-}\frac{1}{2\gamma}\left(\frac{R_{\rm s}}{R_{\rm s,0}}\right)^{2}\left(\frac{l_{\rm mfp,s}}{R_{\rm s,0}}\right)^{-1}\right] \left[\sqrt{\frac{2\gamma}{\pi}}\zeta\exp{\left({-}\frac{\gamma}{2}\zeta^{2}\right)} + \operatorname{erfc}{\left(\sqrt{\frac{\gamma}{2}}\zeta \right)} \right].
\label{eq:phi_final}
\end{equation}

We can now solve Equations~\ref{eq:Rs_R0}, \ref{eq:F1_final}, and \ref{eq:phi_final} numerically to isolate $R_{\rm s}/R_{\rm s,0}$ as a function of $l_{\rm mfp,s}/R_{\rm s,0}$.
\begin{figure}[!htbp]
    \centering
    \includegraphics[width=0.7\textwidth]{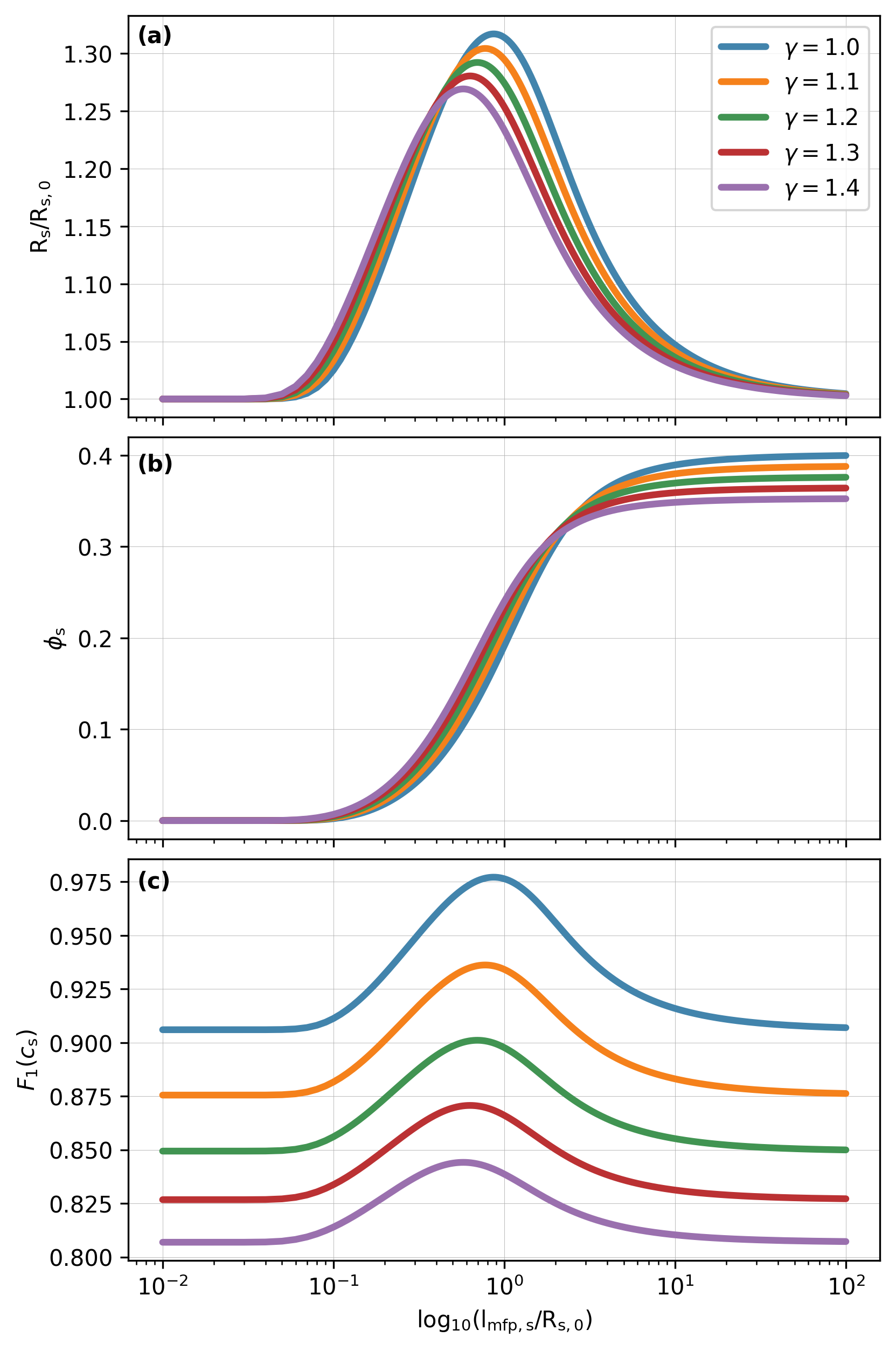}
    \caption{The (a) normalized sonic point radius $R_{\rm s}/R_{\rm s,0}$ (Equation~\ref{eq:Rs_R0}), (b) decoupling fraction at the sonic point $\phi_{\rm s}$ (Equation~\ref{eq:phi_final}), and (c) Maxwellian-tail factor at the sonic point $F_{1}(c_{\rm s,s})$ (Equation~\ref{eq:F1_final}), as a function of the logarithm of $l_{\rm mfp,s}/R_{\rm s,0}$ with $R_{\rm s,0}{=}GM/(2c_{\rm s,s}^{2})$ and $l_{\rm mfp,s}{\equiv}(\sqrt{2}\,n_{\rm s}\sigma)^{-1}$, for $\gamma{=}1,\,1.1,\,1.2,\,1.3$ and 1.4 (blue, orange, green, red, and purple, respectively).}
    \label{fig:plots}
\end{figure}
From Figure~\ref{fig:plots} we see that the sonic point begins at $R_{\rm s}/R_{\rm s,0}{=}1$ when $l_{\rm mfp,s}/R_{\rm s,0}{=}0$, rises to a peak of $R_{\rm s}/R_{\rm s,0}{\approx}1.3$ at $l_{\rm mfp,s}/R_{\rm s,0}{\approx}0.75$, and then returns to $R_{\rm s}/R_{\rm s,0}{\approx}1$ as $l_{\rm mfp,s}/R_{\rm s,0}{\to}\infty$. This behavior follows from Equation~\ref{eq:Rs_R0}. The only term that can shift the sonic point away from $R_{\rm s,0}$ is the coefficient multiplying $R_{\rm s}^{2}$, which is proportional to $\phi_{\rm s}/l_{\rm mfp,s}$. For small $l_{\rm mfp,s}$, $\phi_{\rm s}$ is exponentially suppressed, so the correction is negligible. For large $l_{\rm mfp,s}$, $\phi_{\rm s}$ approaches a finite ceiling set by the Maxwellian tail (Figure~\ref{fig:plots}b), but $1/l_{\rm mfp,s}$ becomes small, again making the correction negligible. As a result, $\phi_{\rm s}/l_{\rm mfp,s}$ peaks at intermediate $l_{\rm mfp,s}$, producing a single, modest outward displacement of $R_{\rm s}$ before the solution returns to the Parker limit. Physically, efficient momentum removal requires both suprathermal particles that can escape and enough collisions to populate that tail, which occurs only at intermediate collisionality. For $R_{\rm s}/R_{\rm s,0}{\approx}1$ and $l_{\rm mfp,s}/R_{\rm s,0}{\to}\infty$, the limiting escape fraction approaches,
\begin{equation}
    \phi_{\rm s,max} = \frac{1}{2} \left[\sqrt{\frac{2\gamma}{\pi}}e^{-\frac{\gamma}{2}} + \operatorname{erfc}{\left(\sqrt{\frac{\gamma}{2}}\right)} \right],
\label{eq:phi_lim}
\end{equation}
suggesting that even at large mean free path, $R_{\rm s}$ retains a finite collisional fraction because $\phi_{\rm s}$ is always bounded from above by $1/2$, and thus at least half of the local collisional flow remains hydrodynamically coupled. The modest displacement of the sonic point radius indicates that the transonic structure of the collisional channel is only weakly modified by particle decoupling. Even at intermediate $l_{\rm mfp,s}$, where the sink strength $\phi_{\rm s}/l_{\rm mfp,s}$ is maximized and the shift in $R_{\rm s}$ is largest, the collisional sonic point radius remains close to the classical Parker value, $R_{\rm s}/R_{\rm s,0}{\lesssim} 1.3$. This indicates that the collisional flow remains close to the Parker solution through the sonic point region, with decoupling mainly mediating the transfer of mass and momentum from the collisional to the collisionless channel. We therefore adopt a two-channel description in which a collisional transonic channel and a collisionless ballistic channel coexist and jointly determine the bulk flow. Section~\ref{sec:two_channel} introduces the two-channel construction used in the remainder of the paper.

\subsection{Two-channel decomposition and how we construct the bulk flow}
\label{sec:two_channel}

In the canonical Parker problem, the wind is treated as a single collisional fluid that conserves mass and momentum \citep{Parker1958,Parker1964,Parker1964b}. Under these assumptions, any steady solution that remains subsonic at all radii asymptotes to a hydrostatic, barometric density profile at large distance \citep[][pages~147$-$151]{walker1977}. Because this tail does not decline rapidly enough as $r{\rightarrow}\infty$, it results in an atmosphere with infinite total mass. In our case, the collisional component is not a closed fluid: mass and momentum are continuously transferred to a collisionless channel as suprathermal particles decouple. The relevant requirement is therefore that the coupled solution has finite mass, not that the flux-weighted bulk flow is everywhere transonic. Moreover, as we showed above, the velocity profile of the collisional channel is only marginally affected by decoupling (equivalently, by changes in $l_{\rm mfp,s}/R_{\rm s}$; see Figure~\ref{fig:plots}a). We therefore approximate the collisional channel as a standard Parker wind when computing its velocity profile, that is, we solve the usual lossless continuity equations for a collisional fluid. We then introduce $\phi(r)$ only to transfer mass from the collisional channel to the collisionless channel. Here and throughout this section, unsubscripted variables refer to the collisional channel, ``nc" for the collisionless channel, and ``tot" for flux-weighted bulk quantities. The resulting Parker expression for the collisional velocity gradient is,
\begin{equation}
    \frac{{\rm d}v}{{\rm d}r} = \frac{2c_{\rm s}^2v}{r^{2}}\frac{\frac{GM}{2 c_{\rm s}^2}-r}{c_{\rm s}^2-v^{2}}.
\label{eq:dv_dr_2}
\end{equation}
The physically relevant transonic solution is selected by the regularity condition at $r{=}R_{\rm s,0}$ and $v{=}c_{\rm s,s}$, yielding the sonic point acceleration (Appendix~\ref{app:acceleration}),
\begin{equation}
    \left.\frac{{\rm d}v}{{\rm d}r}\right|_{\rm s} = \frac{\sqrt{2(5-3\gamma)}-2(\gamma-1)}{\gamma+1} \frac{c_{\rm s,s}}{R_{\rm s,0}}.
\label{eq:dv_dr_c}
\end{equation}

For the collisionless channel, particles move ballistically, so their radial acceleration satisfies $\left.{\rm d}v/{\rm d}t\right|_{\rm nc}{=}{-}GM/r^{2}$. Converting to a radial derivative gives $\left.{\rm d}v/{\rm d}r\right|_{\rm nc}{=}({\rm d}v/{\rm d}t)/({\rm d}r/{\rm d}t){=}{-}GM/(r^{2}v_{\rm nc})$. At a given radius $r$, the collisionless population is not described by a single local Maxwellian: it is a superposition of particles that decoupled at smaller radii from different shifted Maxwellians (set by their local temperature and bulk flow velocity) and then propagated ballistically to $r$ under energy conservation. This superposition does not admit a unique local bulk speed without specifying the full distribution of launch energies, and we therefore adopt $v_{\rm nc}{\approx}v_{\rm esc}(r){=}\sqrt{2GM/r}$, corresponding to threshold escape. Substituting gives,
\begin{equation}
    \left.\frac{{\rm d}v}{{\rm d}r}\right|_{\rm nc} = -\sqrt{\frac{GM}{2r^{3}}}.
\label{eq:dv_dr_nc}
\end{equation}

The bulk, observable velocity is then defined as a flux-weighted average,
\begin{equation}
\langle v\rangle_{\rm tot}(r) = \frac{\rho(r) v(r) + \rho_{\rm nc}(r) v_{\rm nc}(r)}{\rho(r) + \rho_{\rm nc}(r)},
\label{eq:v_tot}
\end{equation}
with the local transfer between channels controlled by the decoupling fraction $\phi(r)$ (Equation~\ref{eq:phi_1}). We obtain $\rho(r)$ by integrating Equation~\ref{eq:mass_conservation}, and $\rho_{\rm nc}(r)$ by integrating the corresponding continuity equation with the same exchange term but opposite sign (i.e., the collisional sink is the collisionless source, i.e., the right-hand side of Equation~\ref{eq:mass_conservation}). In the next section, we show that this construction can produce a maximum in $\langle v\rangle_{\rm tot}$ near $R_{\rm s,0}$ and a breeze-like decline at larger radii as the collisionless channel grows.

\section{Results} 
\label{sec:results}

We solve the collisional and collisionless mass conservation equations by marching outward on a fixed radial grid starting at the lower boundary of the atmosphere, which we set to $10^{-5}\,R_{\rm s}$, and ending at an outer boundary that we treat as a free parameter. Over this domain, we first compute the velocity profiles and their radial gradients. For the collisional channel, we use the Parker limit (Equations~\ref{eq:dv_dr_2} and \ref{eq:dv_dr_c}) by setting the decoupling fraction $\phi{=}0$ when solving for the velocity because decoupling only marginally affects the velocity profile (Figure~\ref{fig:plots}a). For the collisionless channel, particles stream outward without further collisions and therefore move under gravity alone, so the collisionless speed is set to the local escape speed, and its radial gradient is given by Equation~\ref{eq:dv_dr_nc}. 

For the densities, we take the collisional value at the lower boundary from the Parker solution and adjust it until the sonic point density equals unity within a 1\% tolerance. This follows from our nondimensionalization, which scales density by its sonic point value. With the lower boundary value fixed, we integrate the collisional mass conservation equation (Equation~\ref{eq:mass_conservation}) outward using a finite-difference method. The collisionless density is set to zero at the lower boundary, and we integrate the collisionless mass conservation equation outward using the same collisional-to-collisionless transfer term on the right-hand side of Equation~\ref{eq:mass_conservation} but with opposite sign so that mass removed from the collisional component is added to the collisionless component at each radius. 

With the velocities and densities of the collisional and collisionless channels known, we evaluate the bulk mean velocity profile of the wind $\langle v\rangle_{\rm tot}(r)$ (Equation~\ref{eq:v_tot}).
\begin{figure}[!htbp]
    \centering
    \includegraphics[width=1\textwidth]{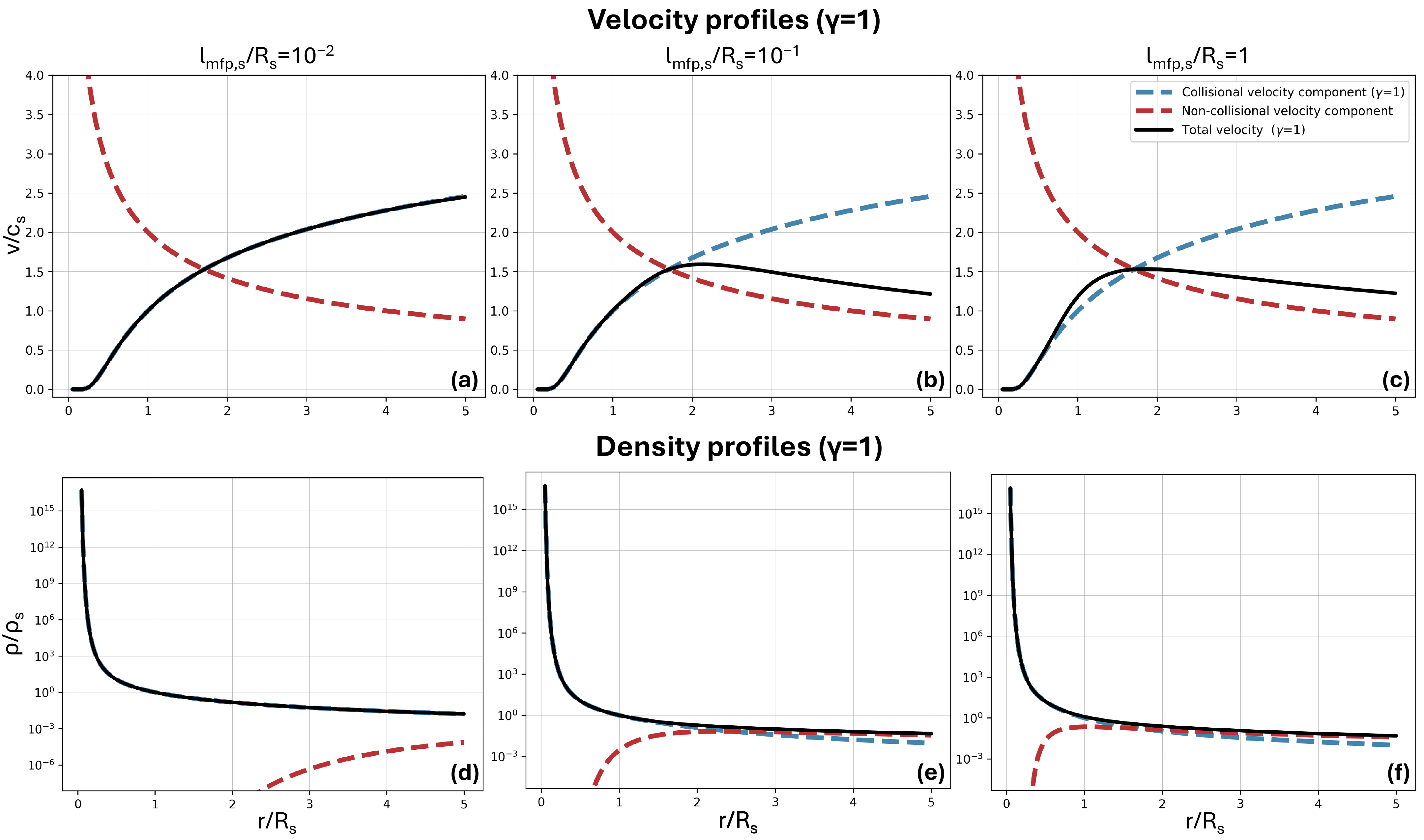}
    \caption{Velocity (top) and density (bottom) profiles for three atmospheres with $\gamma{=}1$ and (left) $l_{\rm mfp,s}/R_{\rm s}{=}0.01$, (middle) $l_{\rm mfp,s}/R_{\rm s}{=}0.1$, and (right) $l_{\rm mfp,s}/R_{\rm s}{=}1$. The blue and red dashed lines show the collisional and collisionless contributions to the flow, and the black solid line shows the total bulk wind.}
    \label{fig:2}
\end{figure}
Figure~\ref{fig:2} summarizes the central result of this work: even when the collisional component satisfies the Parker sonic point regularity condition and continues to accelerate, the flux-weighted bulk velocity can remain breeze-like and decrease at large radii because gradual decoupling transfers mass and momentum to a collisionless channel that decelerates ballistically under gravity. We observe three key behaviors. First, hydrodynamic atmospheres begin with Parker-like acceleration and then evolve into a kinetic molecular outflow. Second, how rapidly the flow deviates from Parker-like behavior with increasing radius depends strongly on the ratio of the mean free path to the sonic point radius $l_{\rm mfp,s}/R_{\rm s}$. This effect is strongest at intermediate temperatures. If the planet is too cold, the atmosphere does not develop a hydrodynamic wind and escape proceeds mainly through Jeans escape \citep{Volkov2011a}. If it is too hot, the scale height is large and the sonic point lies deep in the atmosphere, so the gas remains dense at $R_{\rm s}$ and stays close to the Parker limit \citep[][their Figure~2]{Kubyshkina2018b}. At intermediate temperatures, the sonic point lies higher in the atmosphere while the scale height is small, which lowers the density at $R_{\rm s}$ and enhances decoupling. Third, the onset of kinetic behavior becomes pronounced for $l_{\rm mfp,s}/R_{\rm s}{>}10^{-2}$; at $l_{\rm mfp,s}/R_{\rm s}{=}10^{-2}$ the profile remains close to the Parker solution whereas by $l_{\rm mfp,s}/R_{\rm s}{=}10^{-1}$ it has shifted substantially toward the collisionless limit. This behavior persists for all $\gamma$ values (see Figures~\ref{fig:3} and \ref{fig:4} in the Appendix~\ref{app:sup_figs}), suggesting that a velocity maximum followed by breeze-like, subsonic escape is a natural outcome of partially collisional winds. 

\begin{figure}[!htbp]
    \centering
    \includegraphics[width=\linewidth]{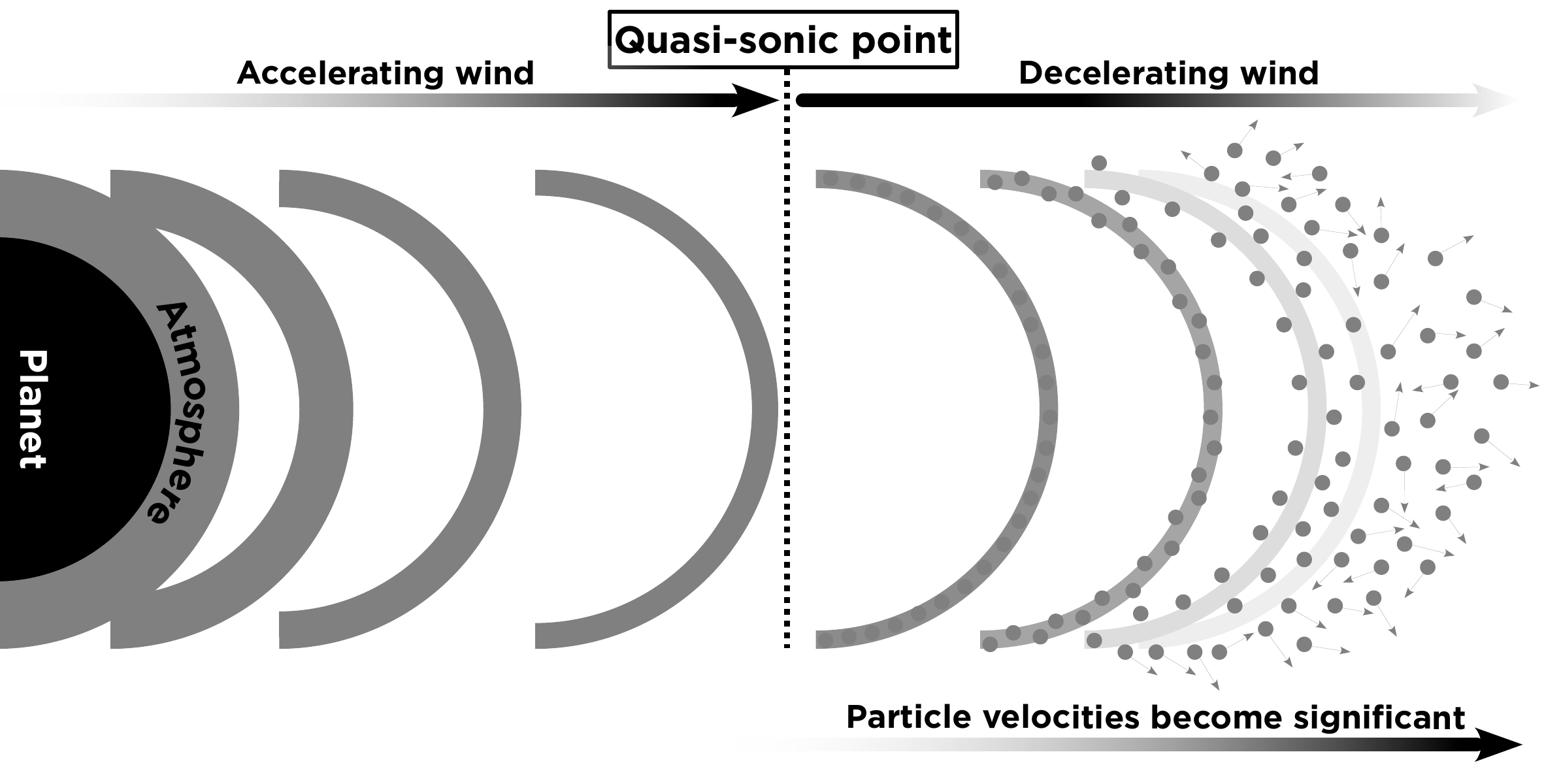}
    \caption{Schematic of the two-channel picture. As the atmosphere expands and rarefies, a fraction of particles decouples from the collisional flow and streams outward ballistically. The collisional channel can remain Parker-like and transonic, while the growing collisionless channel can cause the flux-weighted bulk speed to peak at a quasi-sonic point and decrease at larger radii.}
    \label{fig:transition}
\end{figure}

We define the quasi-sonic point $R_{\rm qs}$ as the radius where the bulk mean velocity $\langle v\rangle_{\rm tot}(r)$ attains its maximum \citep[see][]{Modirrousta2024}. In practice, $R_{\rm qs}$ is determined numerically from the simulated profiles and does not admit a simple closed-form expression because it depends on the full radial structure through $\phi(r)$, which includes an attenuation integral to infinity (Equation~\ref{eq:phi_1}). For partially collisional winds, that is, $l_{\rm mfp,s}/R_{\rm s}$ not too small, $R_{\rm qs}$ lies close to the collisional sonic point radius $R_{\rm s,0}$. In other words, the location where the collisional channel first becomes transonic and hydrodynamic acceleration would otherwise continue is also where particle decoupling becomes significant, so the bulk flow cannot sustain a fully hydrodynamic transonic profile and instead transitions toward a kinetic, subsonic outflow (Figure~\ref{fig:transition}).

This two-channel behavior has an immediate modeling consequence: there is no single radius at which the flow transitions from hydrodynamic to Jeans-like escape. Instead of a binary transition, the collisional and collisionless channels coexist, with mass and momentum continuously transferred from the collisional channel to the collisionless channel as particles decouple. The bulk velocity $\langle v\rangle_{\rm tot}$ is therefore a flux-weighted mixture of a Parker-like collisional channel and a ballistic collisionless channel. The apparent crossing of the two velocity curves does not imply reabsorption because the collisionless channel is defined by the absence of subsequent collisions rather than by a geometrical separation in radius. This can be understood as follows: consider a population of particles embedded in a tenuous region through which a denser, collisional gas passes. Even if both populations occupy the same spatial region, collisions remain stochastic; a particle traversing a distance ${\rm d}r$ through gas of number density $n$ and cross section $\sigma$ experiences an expected number of collisions $\int \sqrt{2} n\sigma\,{\rm d}r$. The probability of suffering no further collisions is therefore $\exp{\left({-}\int \sqrt{2} n\sigma\,{\rm d}r\right)}$, which is bounded by $(0,1)$. The collisionless channel in our model is defined precisely as this subset of particles that undergo no subsequent collisions. It does not correspond to a spatially distinct flow, nor does it imply that decoupled particles are immune to being overtaken by the bulk wind. Particles that do experience further collisions are simply excluded from the collisionless population, whereas those that do not are accounted for by the survival probability, Equation~\ref{eq:P_sur}, in $\phi(r)$. Thus, the two channels represent a decomposition in phase space based on future collision history, not two fluids passing through one another in real space. As a result, the velocity profile can attain a maximum at $R_{\rm qs}$ and then relax to a breeze-like, subsonic outflow even when the exobase lies well above the sonic point (e.g., $l_{\rm mfp,s}/R_{\rm s}{\sim}10^{-1}$; Figure~\ref{fig:2}b). This behavior is not captured by the standard practice of classifying escape as hydrodynamic or Jeans-like by comparing the mean free path to the sonic point radius \citep{Walker1982,Hunten1982,Shizgal1996,Lammer2004,Lammer2013}, but it is consistent with atomistic methods such as Direct Simulation Monte Carlo approaches \citep{Volkov2013a,Mogan2021}.

\section{Discussion} 
\label{sec:discussion}

The central implication of our results is that the bulk velocity profile can be breeze-like even when the collisional component satisfies the Parker sonic point regularity condition. This occurs because gradual decoupling continuously transfers mass and momentum to a collisionless channel, which is subject only to gravity and decelerates with altitude. In what follows, we discuss how a breeze-like bulk profile may affect the interpretation of observations and the limitations and intended scope of the model.

\subsection{On the observability of atmospheric escape}

There have been several detections of escaping hydrogen from exoplanets \citep{Linsky2010,Lecavelier2012,Vidal2013,Rockcliffe2023}. Such studies regularly cite or apply the isothermal Parker wind model to show that external mechanisms are necessary to reproduce the observed outflow velocities of the order ${\sim}100~{\rm km~s^{-1}}$ \citep[e.g.,][]{Owen2023,DosSantos2023}. These velocities are inferred from the wavelength-dependent dimming of the star’s Lyman-$\alpha$ emission during transit, caused by absorption by neutral hydrogen in the planet’s outflow. Interstellar neutral hydrogen along the line of sight also absorbs strongly at Lyman-$\alpha$, making planetary absorption difficult to detect. The planetary outflow, however, has a line-of-sight velocity that Doppler-shifts its absorption relative to the interstellar component, moving it into the observable line wings. Thus, detections primarily trace the column of neutral hydrogen that reaches sufficiently large line-of-sight velocities to produce wing absorption distinct from the interstellar attenuation. An implication of our results is that Lyman-$\alpha$ (non-)detections may encode information about the mass loss regime of the planet. Figure~\ref{fig:2} shows that gradual decoupling can produce a breeze-like flux-weighted bulk velocity that decreases with radius even when the collisional component remains Parker-like. Accordingly, Lyman-$\alpha$ detections are more consistent with outflows that remain collisional and accelerate to large radii, whereas non-detections are consistent with lower-density, partially collisional winds in which the bulk velocity decelerates beyond the quasi-sonic point and the high-velocity neutral column in the line wings is reduced. This could help resolve some puzzling observations such as why some irradiated hydrogen-rich planets show Lyman-$\alpha$ detections of escaping hydrogen \citep{Ehrenreich2015,Lavie2017,dosSantos2020} while others do not \citep{Bourrier2017,Garcia2020,Rockcliffe2021}. This inconsistency even occurs within the same system, where HD~63433~b shows no evidence for a Lyman-$\alpha$ transit while HD~63433~c shows a significant transit \citep{Zhang2022}.

There are, however, limitations to the amount of information that can be retrieved from these observations. First, they provide only a bulk averaged velocity for the outwardly expanding wind across a distance of the order one orbital radian \citep{Owen2023}. Consequently, the gas has probably undergone substantial processing from the interplanetary environment such as through radiation pressure, potentially obscuring information related to the thermodynamic properties of the planetary atmosphere and their role in driving mass loss. Second, observations do not always appear consistent, such as with HD~189733~b, where mass loss was detected in September 2011 but not in April 2010 \citep{Lecavelier2012,Bourrier2013}. This is difficult to explain because it suggests significant atmospheric or environmental variability over geologically negligible timescales. In consideration of the above, it is evident that planetary outflow observations are still in their infancy, and it remains challenging to infer the near-sonic point velocity structure directly from Lyman-$\alpha$ data. Progress will require a larger and more homogeneous sample of systems, as well as repeated observations of the same targets to separate intrinsic variability from line-of-sight and stellar-environment effects.

\subsection{Limitations and intended scope}
\label{sec:limitations}

Our mass loss model is parsimonious by construction. Its purpose is to isolate how collisionality alone shapes the flux-weighted bulk outflow. Real atmospheres include additional physics that we do not attempt to model here: planets rotate and are not perfectly spherically symmetric, winds are intrinsically three-dimensional, and radiative heating and cooling can produce non-monotonic temperature profiles with multiple inversions. Moreover, atmospheres are not monoatomic; composition and chemistry can alter both the thermodynamics and the escape efficiency through dissociation, ionization, and radiative losses \citep{Garcia2007,Murray2009,Kubyshkina2018a}. Our treatment evaluates the collisional velocity profile in the Parker limit while retaining mass exchange through the coupled continuity equations (Equation~\ref{eq:mass_conservation} and its collisionless counterpart). This decomposition is an approximation; its expected error is no larger than the systematic uncertainties that arise in standard numerical treatments of transonic winds, for example from the choice of inner and outer boundary conditions \citep[][their Section~2.2 and Appendix~A]{Murray2009} or from the artificial dissipation used to regularize the sonic transition \citep{Garcia2007}. 

Partially collisional winds are most relevant for warm to hot super-Earths and sub-Neptunes, where the upper atmosphere can become rarefied while remaining only weakly ionized. By contrast, gas-giant winds can remain effectively fluid if ionization produces a magnetized ion population that maintains long-range correlations and couples to neutrals through ion-neutral collisions. Accordingly, we next assess whether ion-mediated coupling can maintain fluid-like behavior even when neutral-neutral collisions are rare. In this case, magnetized ions provide long-range correlations set by the ion Larmor radius, and frequent ion-neutral collisions transmit ion momentum to neutrals, maintaining bulk fluid behavior. In hot Jupiters, where thermospheric temperatures can readily exceed 10,000~K \citep{Garcia2019,Garcia2023,Huang2023}, ionization may indeed allow for global correlation. As we show below, however, for super-Earths and sub-Neptunes, which have significantly lower thermospheric temperatures \citep[${\lesssim} 4000~{\rm K}$;][]{Lammer2013,Kubyshkina2018a,Caldiroli2021}, this does not rescue collisional behavior for the neutral component in the upper atmospheres.

For the ionized component, the relevant correlation scale is the ion Larmor radius, $l_{\rm i} {=} \mu_{\rm i} v_{\rm i}/(q_{\rm i} B)$, where $\mu_{\rm i}$, $v_{\rm i}$, $q_{\rm i}$, and $B$ are the ion mass, velocity, charge, and the ambient magnetic field strength. Requiring the Larmor radius to be smaller than the sonic point radius and solving for $B$ shows that, for standard parameter choices, a magnetic field ten orders of magnitude weaker than Earth’s field is sufficient to keep the ionized species correlated. Ion coupling alone, however, does not guarantee that the neutrals are correlated. For the neutral component to behave as a fluid, ion-neutral collisions must be frequent enough for the magnetized ions to exchange momentum with neutrals. This can be approximately evaluated by comparing the effective mean free path between ions and neutrals, $\langle l\rangle {\sim} v_{\rm n}/\left(n_{\rm i}\langle \sigma v_{\rm rel}\rangle\right)$ \citep[][Section~2.3]{Ballester2018}, where $v_{\rm n}$ is the characteristic thermal speed of a neutral particle and $v_{\rm rel}$ is the relative speed between ions and neutrals that enters the thermally averaged rate coefficient $\langle\sigma v_{\rm rel}\rangle$. Letting $v_{\rm n}{\sim}c_{\rm s}$ and recognizing that $\langle \sigma v\rangle{\sim}3{\times}10^{-15}~{\rm m^{3}~s^{-1}}$ \citep{Pinto2008} one finds that $n_{\rm i}{\gtrsim} 10^{12}\,{\rm m}^{-3}$ is required for ions to couple neutrals over a scale shorter than the sonic point radius. This threshold is many orders of magnitude larger than the densities expected in the upper atmosphere of rocky exoplanets with primordial atmospheres \citep{Sekiya1980,Sekiya1981,Kubyshkina2018a,Caldiroli2021}. Moreover, at the relatively low thermospheric temperatures expected for super-Earths and sub-Neptunes \citep[${\lesssim} 4000~{\rm K}$;][]{Lammer2013,Kubyshkina2018a,Caldiroli2021}, the ionization fraction is small, so $n_{\rm i}$ is an even smaller subset of an already rarefied gas, indicating that ion-neutral coupling does not rescue fluid behavior for the neutral component.

\section{Conclusion}
\label{sec:conclusion}

Our paper shows that purely hydrodynamic descriptions of atmospheric escape can be inaccurate because a finite fraction of particles continuously decouples from the collisional flow and streams outward collisionlessly. Even when the mean free path at the sonic point is a small fraction of the sonic point radius, suprathermal particles populate a collisionless channel and remove mass and momentum from the collisional component. As a consequence, single-component hydrodynamic models can overestimate the bulk speed at large radii and predict a transonic bulk solution even when the total escape remains subsonic and is dominated by collisionless transport. A breeze-like bulk profile should therefore not be interpreted as a failure to realize a Parker-type transonic collisional solution; it arises naturally once the escaping flux is allowed to transition continuously from collisional to collisionless transport. We find that the collisional and collisionless components coexist over an extended region and that the bulk outflow departs strongly from a Parker solution for $l_{\rm mfp,s}/R_{\rm s}{>}10^{-2}$, a range that is often treated as fully hydrodynamic: the collisional channel still passes through its sonic point, but the bulk reaches a maximum at the quasi-sonic point and then declines outward.

\section*{Acknowledgements}
We thank the referee for their constructive comments that improved the manuscript, in particular by bringing to our attention the link between our model and Lyman-$\alpha$ observability through the distinction between accelerating and decelerating bulk outflows.

\section*{Appendix}
\setcounter{equation}{0}
\setcounter{figure}{3}
\setcounter{section}{0}

\renewcommand{\thesection}{A\arabic{section}}
\renewcommand{\theequation}{\thesection.\arabic{equation}}
\renewcommand{\thefigure}{A\arabic{figure}}

\section{Deriving the mean free path}
\label{app:mfp}

Consider a particle with collision cross section $\sigma$ moving through a monoatomic gas of number density $n$. Collisions are controlled by how quickly the particle and potential partners approach one another, so the relevant speed for counting encounters is the relative speed $v_{\rm rel}$. Over a time interval $\Delta t$, the particle sweeps out the volume $\Delta V {=} \sigma v_{\rm rel}\Delta t$ so that the expected number of collision partners within that volume is $n\Delta V$. Therefore, the collision rate is $\nu {=}n\Delta V/\Delta t{=} n\sigma v_{\rm rel}$ and the mean time between collisions is $\Delta t_{\rm col} {=} \nu^{-1} {=} \left(n\sigma v_{\rm rel}\right)^{-1}$. The mean free path is defined as the mean distance traveled by the particle between collisions, which is set by the particle’s own speed $v$ and not by the closing speed with a partner $v_{\rm rel}$.  In general $v$ and $v_{\rm rel}$ are not the same because the other gas molecules are also moving, so the rate at which collision partners are encountered depends on their motion as well, whereas the distance the particle advances in space depends only on its own motion. Therefore the distance traveled between collisions is $l_{\rm mfp} {=} v\Delta t_{\rm col}$, giving $l_{\rm mfp} {=} v/\left(n\sigma v_{\rm rel}\right)$. To estimate $v_{\rm rel}$, we consider two uncorrelated particles with velocities $\vec{v}_{1}$ and $\vec{v}_{2}$. Their relative speed is $v_{\rm rel} {=} \sqrt{\left(\vec{v}_{2}{-}\vec{v}_{1}\right)\cdot\left(\vec{v}_{2}{-}\vec{v}_{1}\right)} {=} \sqrt{v_{1}^{2}{+}v_{2}^{2}{-}2\vec{v}_{1}{\cdot}\vec{v}_{2}}$. For an isotropic distribution with uncorrelated directions, $\langle \vec{v}_{1}{\cdot}\vec{v}_{2}\rangle {=} 0$, and for equal characteristic speeds $v_{1}{=}v_{2}{=}v$ this gives $v_{\rm rel} {=} \sqrt{2}v$. Substituting back yields $l_{\rm mfp} {=} \left(\sqrt{2}n\sigma\right)^{-1}$.

\section{Deriving the mass sink}
\label{app:mass_sink}

To understand why the product of the number density of escaping particles at radius $r$, $n\phi$, the inverse mean free path, $l_{\rm mfp}^{-1}(r){=}\sqrt{2}n(r)\sigma$, and the outward radial speed with which such particles leave the collisional fluid, $v{+}\bar{v}_{\rm esc}$, gives the local mass transfer rate per unit volume, consider a thin radial shell of thickness ${\rm d}r$ at radius $r$. The particles eligible to leave the collisional channel have local number density $n\phi$. In a time interval ${\rm d}t$, those particles travel outward by ${\rm d}r{=}(v{+}\bar{v}_{\rm esc}){\rm d}t$. For ${\rm d}r{\ll} l_{\rm mfp}$, the probability that any one particle experiences a collision while crossing the layer is ${\rm d}r/l_{\rm mfp}$. Therefore the expected number of particles per unit volume that undergo such a collision during ${\rm d}t$ is $n\phi{\rm d}r/l_{\rm mfp} {=} (v{+}\bar{v}_{\rm esc})n\phi{\rm d}t/l_{\rm mfp}$. Dividing by ${\rm d}t$ and multiplying by $\bar\mu$ gives the corresponding mass sink per unit volume $\sqrt{2}n^{2}\sigma \phi (v{+}\bar v_{\rm esc})$.

\section{Deriving the acceleration at the sonic point}
\label{app:acceleration}

Differentiating the numerator and denominator of Equation~\ref{eq:dv_dr_2} and letting $r{=}R_{\rm s}$ and $v{=}c_{\rm s}$, 
\begin{equation}
    \left.\frac{{\rm d}v}{{\rm d}r}\right|_{\rm c} = \frac{c_{\rm s,s}}{R^{2}_{\rm s}} \frac{R_{\rm s}\frac{{\rm d}\ln{T}}{{\rm d}r}+1}{\frac{1}{c_{\rm s,s}}\frac{{\rm d}v}{{\rm d}r}-\frac{1}{2}\frac{{\rm d}\ln{T}}{{\rm d}r}}.
\label{eq:app_1}
\end{equation}
The temperature derivative is found by differentiating the polytropic relation $T{\propto}P^{(\gamma-1)/\gamma}$ with respect to distance, and combining with Equation~\ref{eq:mom_conservation} set at $\phi(r){=}0$,
\begin{equation}
    \frac{{\rm d}\ln{T}}{{\rm d}r} = -2\left(\gamma-1\right)\left(\frac{1}{R_{\rm s}} + \frac{1}{2c_{\rm s,s}}\frac{{\rm d}v}{{\rm d}r} \right).
\label{eq:app_2}
\end{equation}
Combining Equations~\ref{eq:app_1} and \ref{eq:app_2} yields,
\begin{equation}
    \frac{\gamma+1}{4(\gamma-1)}\frac{R_{\rm s}}{c_{\rm s,s}} \left(\frac{{\rm d}v}{{\rm d}r}\right)^{2} + \frac{{\rm d}v}{{\rm d}r} + \frac{2\gamma-3}{2(\gamma-1)}\frac{c_{\rm s,s}}{R_{\rm s}} = 0,
\label{eq:app_3}
\end{equation}
which can be solved with the quadratic formula to give Equation~\ref{eq:dv_dr_c}.

\section{Supplementary figures}
\label{app:sup_figs}

\begin{figure}[!htbp]
    \centering
    \includegraphics[width=1\textwidth]{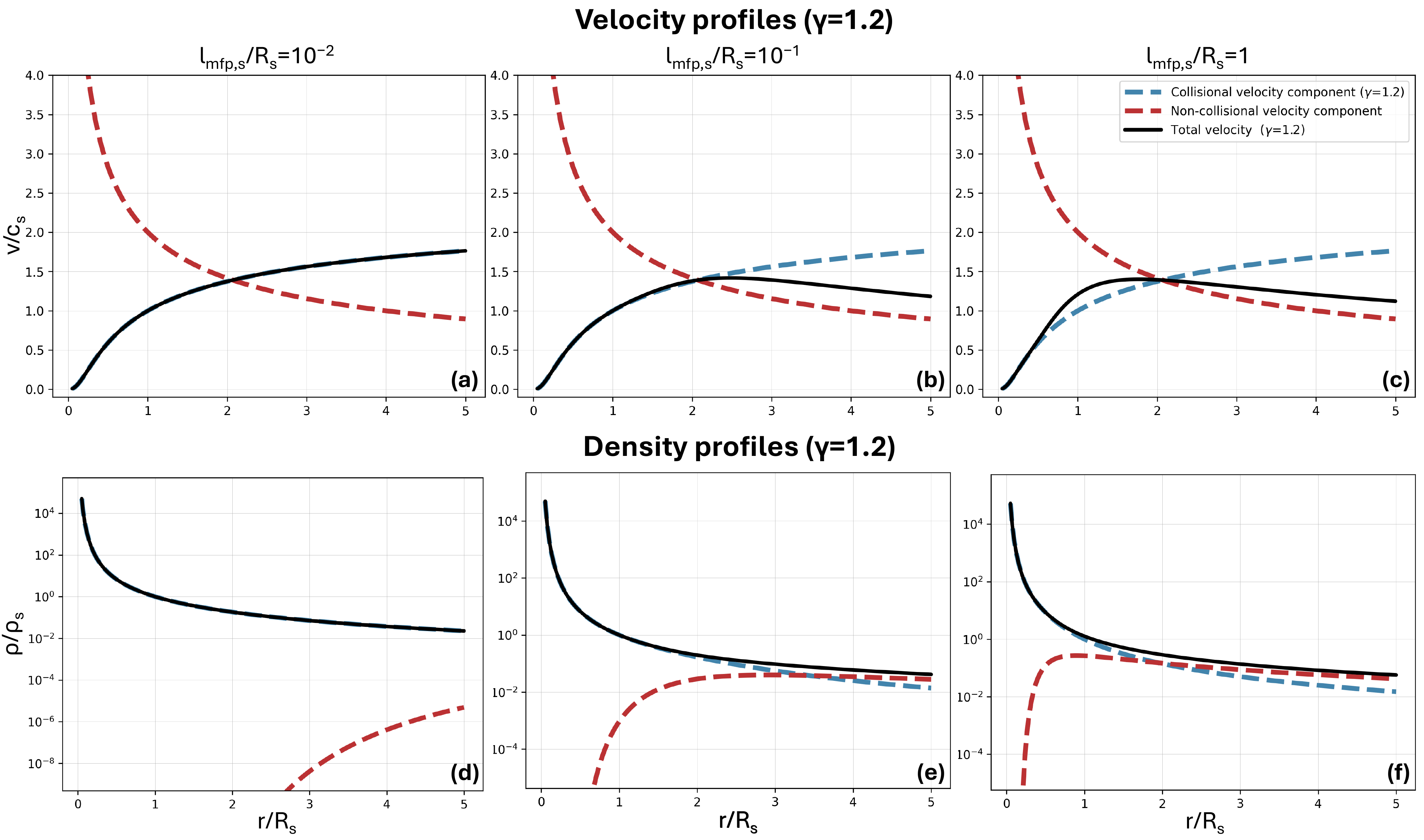}
    \caption{Velocity (top) and density (bottom) profiles for three atmospheres with $\gamma{=}1.2$ and (left) $l_{\rm mfp,s}/R_{\rm s}{=}0.01$, (middle) $l_{\rm mfp,s}/R_{\rm s}{=}0.1$, and (right) $l_{\rm mfp,s}/R_{\rm s}{=}1$. The blue and red dashed lines show the collisional and collisionless contributions to the flow, and the black solid line shows the total bulk wind.}
    \label{fig:3}
\end{figure}

\begin{figure}[!htbp]
    \centering
    \includegraphics[width=1\textwidth]{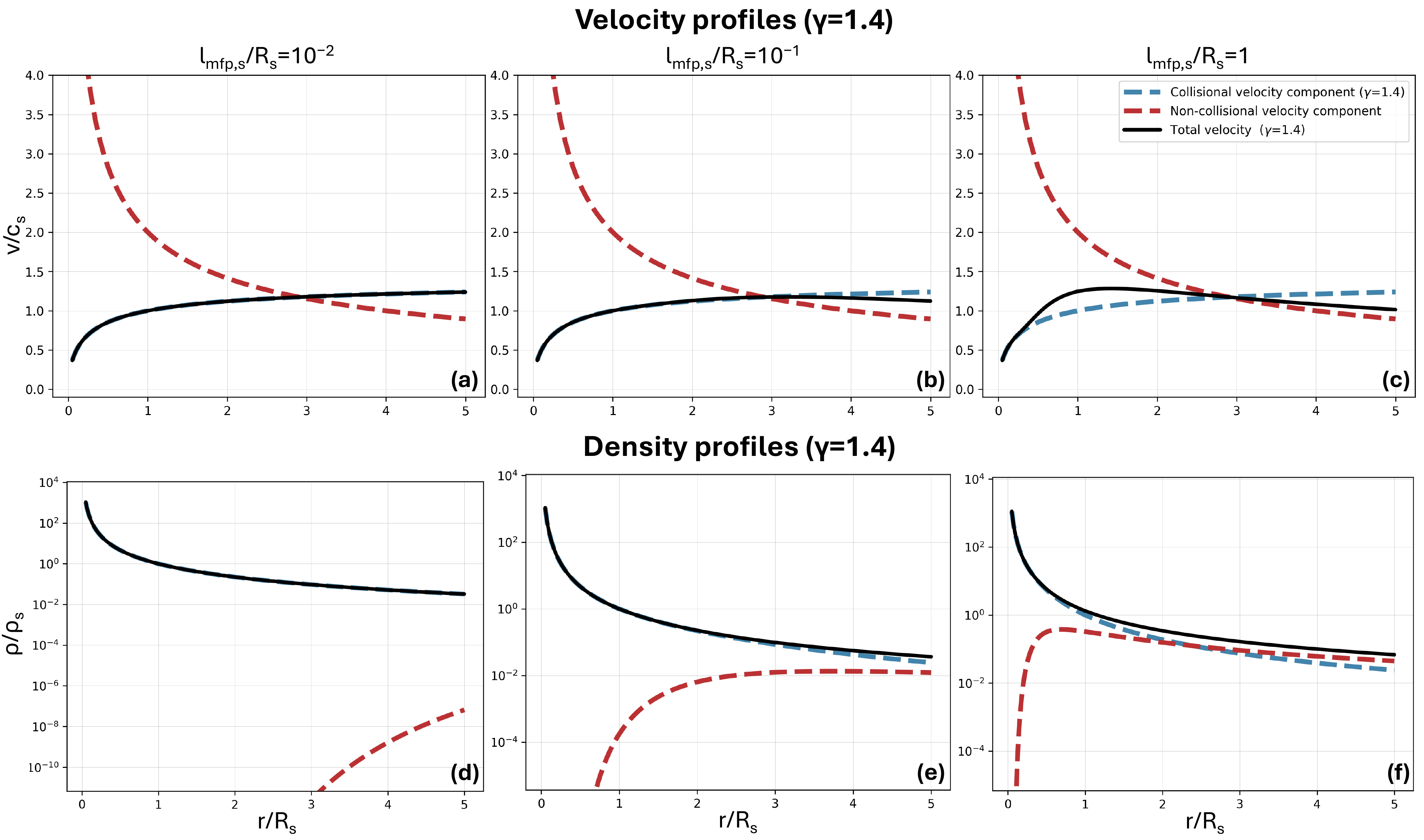}
    \caption{Velocity (top) and density (bottom) profiles for three atmospheres with $\gamma{=}1.4$ and (left) $l_{\rm mfp,s}/R_{\rm s}{=}0.01$, (middle) $l_{\rm mfp,s}/R_{\rm s}{=}0.1$, and (right) $l_{\rm mfp,s}/R_{\rm s}{=}1$. The blue and red dashed lines show the collisional and collisionless contributions to the flow, and the black solid line shows the total bulk wind.}
    \label{fig:4}
\end{figure}

Figures~\ref{fig:3} and \ref{fig:4} show velocity and density profiles for atmospheres with polytropic indices $\gamma{=}1.2$ and $1.4$ for different values of the ratio of the mean free path at the sonic point to the sonic point radius, $l_{\rm mfp,s}/R_{\rm s}$. These profiles illustrate how varying $l_{\rm mfp,s}/R_{\rm s}$ changes the relative contributions of the collisional and collisionless components of the flow.

\bibliography{sample701}{}
\bibliographystyle{aasjournal}

\end{document}